# Reliable isomorphic physics problem generation with large language models

Xian Wu and Lindim Ismaili
*Department of Physics, University of Connecticut, Storrs, CT, 06269 USA*

This study presents an AI-powered system for generating isomorphic physics problems using large language model (LLM)-based agent workflows. The system is designed around three practical goals: preserving the same conceptual and problem-solving structure as the original problems, varying construct-irrelevant features such as scenarios and numerical values, and producing questions that are directly usable without expert revision. The workflow combines prompt chaining, agent-based verification, and automated LaTeX compilation within a publicly accessible website hosted on a Raspberry Pi. To evaluate the system, we developed an eight-item rubric and tested the system using 13 multiple-choice questions from a calculus-based introductory Newtonian mechanics course. The evaluation results showed that 89% of the questions generated were rated as fully specified and directly usable. However, the system also showed limitations. The results suggest that LLM-based systems have significant potential for reliable instructional problem generation while also highlighting important challenges for future development.

# I. INTRODUCTION

Recent advances in large language models (LLMs), particularly OpenAI's Chat Generative Pre-Trained Transformer (ChatGPT) [1] have rapidly transformed educational technology and created new opportunities for personalized, adaptive, and interactive learning environments. A search of the Education Resources Information Center (ERIC) database conducted on May 9, 2026, using the search string "large language model OR llm OR chatgpt" returned 1,844 peer-reviewed articles. Only 14 of these articles were published before 2023, with the earliest article in the database published in 2007; by contrast, 925 articles were published in 2025 alone. Many recent systematic reviews have reported that LLM-based educational systems can support adaptive learning pathways, automated assessment, and real-time learning assistance, while also improving student engagement and learning efficiency in some contexts [2–5].

Despite these promising developments, the current literature consistently highlights substantial limitations and unresolved concerns surrounding the implementation of LLMs in education. One major challenge is the reliability of AI-generated outputs, particularly hallucinations that produce inaccurate or misleading educational explanations [2,3]. Beyond factual reliability, several studies emphasize that current LLM-generated educational support often lacks the pedagogical depth and instructional appropriateness of human educators [2]. Researchers have also expressed concern that excessive dependence on AI-generated assistance may weaken students' critical thinking and independent problem-solving abilities over time [2–5]. Consequently, many researchers conclude that LLMs should be used as supplemental tools that require oversight from educational experts.

The current project aims to make progress in this area by examining whether LLMs can generate isomorphic physics problems based on uploaded source problems without human revision. Physics instructors often need to create multiple versions of problems, such as changing numerical values across semesters, creating multiple versions of the same test, or modifying homework and in-class examples into exam problems. Current online homework systems offered by major publishers usually rely on built-in algorithms to randomize numerical values within existing problems. However, these systems generally cannot generate problems with new scenarios beyond existing script. It would therefore be valuable if LLMs could generate new problems that preserve the structure of original problems while varying contextual and numerical features, with minimal or no expert revision. The research question guiding this project is: **To what extent can problems generated by LLMs be directly used as instructional content without expert revision?**

# II. ISOMORPHIC PROBLEM GENERATION

## A. Isomorphic problems

Isomorphic problems are often used to support and assess student learning [6,7]. A central issue, however, is what makes two problems isomorphic. Early definitions describe a problem as a task that presents a gap between the initial information and a desired goal, where the path to the goal is not immediately obvious and must be discovered through a series of deliberate operations that reorganize the available information [8,9]. Based on this understanding, Simon and Hayes [10] defined isomorphic problems as problems whose successive situations can be mapped in a one-to-one fashion. This definition does not fully match physics problem solving, because people may use different solution approaches for the same problem. Physics problems often combine conceptual understanding with mathematical manipulation, making it unrealistic to expect two loosely similar problems to be solved through procedures that are mapped in a strict one-to-one fashion. More recently, researchers have tended to use more practical definitions of isomorphic problems. In general, two problems can be considered isomorphic if they require the same core concepts or principles to solve them [6,11].

It is important to note that this interpretation largely reflects an expert view. From the "knowledge in pieces" perspective [12,13], novices' use of physics knowledge is highly sensitive to surface features and may differ substantially from experts' organization of the same knowledge. Therefore, problems that are isomorphic from an expert perspective may not necessarily be perceived as isomorphic by novice students.

## B. Problem generation by LLM

There are two major approaches to using LLMs to generate problems for instructional or assessment purposes. The first approach is to generate new problems based on lecture notes, learning objectives, and/or knowledge structures. Shu et al [14] extracted learning objectives from lecture notes in undergraduate engineering courses and then used LLMs to generate multiple-choice questions (MCQs) based on those learning objectives. They evaluated the generated MCQs and reported that more than 90% of the questions were usable. However, they rated generated MCQs from 1 to 5 for each criterion without providing detailed information about what each numerical rating meant. For example, one of their rating criteria was "distractor option quality," but the article did not clearly define what constitutes an excellent distractor. Zhu et al [15] proposed an automatic MCQ generation system for K–12 science classes based on a knowledge network built by educational experts. They evaluated the semantic similarity and difficulty of the generated questions but did not directly evaluate the overall quality or instructional usability of the questions. The second approach generates new problems from existing problems. Liu et al [16] generated a question bank for an introductory Newtonian mechanics course by engineering prompt chains. In their study, both students and local LLMs were used to evaluate the similarity of the generated isomorphic questions and the quality of all generated questions. However, the authors only briefly mentioned that numerical and language issues were detected by students and local LLMs. They did not provide quantitative data indicating the extent to which the generated questions could be used directly without revision.

FIG. 1. A screenshot of the website showing the public interface and workflow.

### C. Evaluation rubric

Because there is no commonly accepted definition of "isomorphic problems," we designed our system around three goals: generated problems should be conceptually the same as the originals, different in construct-irrelevant features, and directly usable without revision. Hereafter, we refer to these goals as "same, different, and directly usable."

The "same" requirement means that generated problems should test the same physics concepts as the original problems and, as much as possible, require similar problem-solving approaches. For open-ended problems, this means preserving the conceptual structure while changing the scenario and, when applicable, modifying numerical values. For MCQs, this additionally means preserving the logical structure of the distractors, because many MCQs use common student misconceptions to construct plausible incorrect options. The "different" requirement means that generated problems should use different scenarios and, when applicable, different numerical values. The new numerical values should also remain physically reasonable in the new context. The "directly usable" requirement means that generated problems must be solvable as written and free of physics errors, numerical errors, typographical errors, grammatical errors, and ambiguous wording.

Based on these design principles, we developed an evaluation rubric with eight items. The first two rubric items address the "same" goal. Because distractors in MCQs are often based on common student misconceptions, we evaluated whether generated MCQs preserved not only the same physical content but also the logical structure of the distractors. The next two rubric items address the "different" goal by evaluating whether the generated problems use different contextual and numerical information. The final four rubric items address whether the generated problems are error-free and usable without revision. Each rubric item was evaluated as either "yes" or "no" to minimize ambiguity. The full evaluation rubric is shown in TABLE I.

## III. FUNCTIONS AND EVALUATION OF THE SYSTEM

### A. AI-powered isomorphic question generation and website hosting

Our workflow uses prompt chaining, structured agent-to-agent data passing, and containerized tool use to generate isomorphic physics problems from uploaded PDF files, DOCX files, or text inputs. Before the agent workflow begins, the backend uploads the source files to OpenAI, creates a shared OpenAI container for the run, and initializes that container with document-processing and LaTeX dependencies. This shared runtime is then made available to the agents through tool calls, including Code Interpreter access for file parsing and computational verification, as well as a custom LaTeX compilation tool for producing final PDF artifacts. These operations are carried out primarily through Python.

The agent chain proceeds through several specialized stages. The Normalizer agent converts the user request into a structured generation configuration. The Extractor agent reads the uploaded materials and converts source problems, metadata, mathematical notation, answer choices, and problem types into a schema-conformant text file. The Verifier agent solves or checks the extracted items and adds answer explanations. The Generator agent creates new isomorphic problems that preserve the same physics concept, reasoning path, part structure, and distractor logic while varying construct-irrelevant features such as context, wording, numerical values, units, and answer order. Finally, the Latexifier agent formats the generated text file output into a complete LaTeX document and calls the containerized PDF compilation function. The workflow is designed so that each agent passes structured information to the next stage of the system. This process preserves problem structure, answer choices, and formatting across the workflow while reducing ambiguity and enabling automated verification and retry logic. These design features are important for generating problems that satisfy the rubric criteria of being "same, different, and directly usable."

The website provides a public interface for this workflow through a Raspberry Pi server that hosts both the frontend and backend of the system. The project code is stored in a GitHub repository and deployed to a Raspberry Pi, where the web application runs continuously. The links of the website and the GitHub repository are included in the appendix. Users access the website through a domain registered and managed through Cloudflare. Cloudflare connects the public domain to the Raspberry Pi server and manages Domain Name System (DNS) settings, security, and web traffic routing. After a user uploads source materials, the system sends the content to LLM agents hosted in an OpenAI runtime through API requests. The agents analyze the uploaded problems, generate isomorphic versions, verify the generated output, and format the final results in LaTeX. The LaTeX code is then compiled into a PDF file, returned to the Raspberry Pi server, and served to the user as a downloadable file or website preview. Fig. 1 shows a screenshot of our website running a testing trial.

### B. Isomorphic multiple-choice question generation and evaluation

We tested the system with 13 MCQs written by one of the authors. These MCQs were designed for a calculus-based introductory Newtonian mechanics course and had been used as quiz questions for three years. The set included a range of formats: four conceptual questions, four numerical questions with abstract contexts (e.g., "if the displacement of an object is given by the function…"), and five numerical questions with contextual information (e.g., "The raven is sent from Castle Black, which is at the origin, observing the white walkers at the coordinates…").

We uploaded a PDF file containing all 13 questions to the website and asked the system to generate LaTeX code that could be converted into a PDF file containing 13 new questions in one trial. We ran 20 trials in total. One of the 20 trials failed to generate LaTeX code as the output. The remaining 19 successful trials produced 247 generated questions in total (13 questions × 19 trials). Example PDFs of the original and generated MCQs are provided in the appendix.

We evaluated all 247 generated questions and tallied the results for each rubric item across the 19 trials. Two authors

Q1. Which of the following statements is correct?

A. **We can treat a bicycle wheel as a point mass when studying how the spokes flex as it rolls.**
This is incorrect because spoke flex depends on the wheel's size and shape, so a point-mass model throws away the needed geometry.

B. **We can reasonably treat a delivery van as a point mass when estimating how far it travels across town in 15 minutes.**
For city-scale motion, the van's size is tiny compared with the travel distance, so a point-mass approximation captures its position change well.

C. **We are not allowed to assume motion in a vacuum even if we want to neglect air resistance in a first-pass calculation.**
This is incorrect because "ignore air resistance" is essentially the same modeling step as treating the situation like a vacuum for that effect.

D. **None of the above is correct.**
This is incorrect because one of the statements (about using a point-mass model over large distances) is a valid approximation.

Q4. A scout drone departs from a base located at the origin and detects a signal at position $\vec{r} = 4\,\hat{\imath} + 7\,\hat{\jmath}$, where distances are measured in megameters. Which of the following statements is correct?

A. **The distance between the signal and the base is 11 megameters.**
This is incorrect because 11 is the sum of components, not the Euclidean distance.

B. **The distance between the signal and the base is 8 megameters.**
This is incorrect because the actual distance is $\sqrt{4^2 + 7^2} = \sqrt{65} \approx 8.06$, not exactly 8.

C. **The distance between the signal and the base is 7 megameters.**
This is incorrect because 7 is just the y-component, not the magnitude of the position vector.

D. **None of the above is correct.**
This is correct because the distance is $\sqrt{65} \approx 8.06$ megameters, which is not 11, 8, or 7.

FIG. 2. Two problematic question examples.

independently evaluated the generated questions using the evaluation rubric. They then compared notes and discussed all discrepancies to reach a final agreement. TABLE I presents the evaluation results agreed upon by the two authors.

Based on the evaluation results, 89% of the generated questions were fully specified and solvable as written, suggesting that most of the LLM-generated questions can be used directly as instructional content without expert revision. The top panel of Fig. 2 shows a generated question rated as having clerical errors and not being solvable as written because it used the phrase "first-pass" in the question statement. This phrase is not a well-defined scientific term and may lead students to interpret the answer choices in different ways.

One major limitation of the system is that it did not consistently preserve distractor logic. The lower panel of Fig. 2 shows a representative example. This generated question was rated as failing to preserve the logical structure of the distractors and as not solvable as written. The original question used the x- and y-components as two distractors, but the generated question failed to preserve this structure. The question was also unsolvable because option B was actually the correct answer, although the system marked option D, "None of the above is correct," as correct by exaggerating the difference between exact and rounded results. In addition, the system did not perform well in changing the scenario for questions with abstract contexts. Because the generated questions closely followed the style of the original questions, the system appeared to have difficulty modifying abstract terms such as "object" and "displacement." We therefore rated these questions as failing to use scenarios different from the original questions when they only randomized the numerical values.

TABLE I. Evaluation results for generated MCQs.

| Evaluation Rubric | Overall | Conceptual | Abstract Context | Numerical |
|---|---|---|---|---|
| Does the generated question preserve the original physical content? | 95% | 100% | 100% | 87% |
| Does the generated question preserve the logical structure of the distractors? | 59% | 68% | 67% | 47% |
| Does the generated question use the scenario different from the original one? | 69% | 99% | 5% | 97% |
| If the original question is numerical, does the generated question appropriately modify numerical values? | 97% | NA | 100% | 95% |
| Is the generated question free of physics errors or misconceptions? | 94% | 80% | 100% | 100% |
| Do all numerical values have physically reasonable magnitudes and units? | 100% | NA | 100% | 100% |
| Is the generated question free of clerical errors (e.g., typos, missing symbols, ambiguous wording)? | 96% | 97% | 100% | 92% |
| Is the generated question fully specified and solvable as written? | 89% | 76% | 100% | 92% |

### C. Work-in-progress functions

Because our system does not impose requirements on a problem's topic, content, or format, it can potentially generate many types of problems across different domains. In this section, we discuss several functions we have tested but not yet rigorously evaluated. Additional generated examples are included in the appendix.

#### *1. Recognizing and generating MCQs based on a shared scenario*

Multiple MCQs are often grouped around the same scenario while testing different aspects of one or more concepts. Our system can recognize this structure and generate new questions that preserve the same shared-scenario structure. We tested this function using the Quantum Mechanics Concept Assessment (QMCA) [17], which includes several instances of multiple questions based on a shared scenario. In this preliminary test, the generated questions preserved the same structure.

#### *2. Generating context-rich problems*

The system can also be used to generate new open-ended problems. One potential application is the generation of context-rich problems [18]. Context-rich problems are valuable for facilitating student discussion and supporting problem-solving skills, but they can be time-consuming to design. It would also be useful to leverage the language-processing capabilities of LLMs to generate context-rich problems with specific thematic focuses, such as engineering-focused, life-sciences-focused, or astronomy-focused scenarios. We added this function to the system and conducted a preliminary test. The results were not satisfactory. The generated problems failed to mimic the story-like style of the original context-rich problems. Changing the thematic focus mainly caused the system to replace objects and surface-level cover stories rather than meaningfully redesign the scenario.

## IV. CONCLUSIONS AND FURTHER WORKING DIRECTIONS

In this study, we developed and evaluated an AI-powered system for generating isomorphic physics problems using LLM-based agent workflows. The system was designed around three practical goals for generated problems: they should preserve the same conceptual and problem-solving structure as the original problems, differ in construct-irrelevant features such as scenarios and numerical values, and be directly usable without expert revision. The evaluation results showed that most generated questions preserved the intended physics content and were solvable as written.

Future work will focus on further improving the reliability of the system, particularly in preserving distractor logic for MCQs and generating context-rich problems that better match the narrative depth and contextual style of the original problems. More broadly, the long-term goal of this project is to develop an AI-powered tutoring system in which reliable problem generation serves as a core component for providing students with timely and personalized instructional content. Such a system could adapt generated problems according to students' backgrounds, interests, and learning needs, including addressing specific misconceptions, strengthening mathematical manipulation skills, and supporting different learning pathways.

## APPENDIX

The public-accessible website of the problem-generation system is available at https://uconn.wuxian123.com/. The project code is hosted on GitHub at https://github.com/lindimismaili/UConn-IsomorphicQuestionLLM. PDF files of the original and generated questions are available in a Google Drive folder: https://drive.google.com/drive/folders/1pZlZy1FIY1cpa4cod89TqDIXXgXWypd2?usp=sharing.